\documentclass[a4paper,11pt]{article}
\usepackage{pos}

\title{Galaxy Cluster simulations with a spectral Cosmic Ray model - "Wrong Way" Radio Relics}
 \ShortTitle{"Wrong Way" Radio Relics}

\author*[a, b]{Ludwig M. Böss}
\author[c]{Ulrich P. Steinwandel}
\author[a,d]{Klaus Dolag}

\affiliation[a]{University Observatory Munich, Scheinerstr. 1, D-81679 Munich, Germany}
\affiliation[b]{Excellence Cluster ORIGINS, Boltzmannstr. 2, D-85748, Garching, Germany}
\affiliation[c]{Center for Computational Astrophysics, Flatiron Institute, 162 5th Avenue, New York, NY 10010}
\affiliation[d]{Max Planck Institute for Astrophysics, Karl-Schwarzschildstr. 1, D-85748, Garching, Germany}

\emailAdd{lboess@usm.lmu.de}

\abstract{Non-thermal emission from relativistic electrons gives insight into the strength and morphology of intra-cluster magnetic fields, as well as providing powerful tracers of structure formation shocks. Emission caused by Cosmic Ray (CR) protons on the other hand still challenges current observations and is therefore testing models of proton acceleration at intra-cluster shocks. Large-scale simulations including the effects of CRs have been difficult to achieve and have been mainly reduced to simulating an overall energy budget, or tracing CR populations in post-processing of simulation output and has often been done for either protons or electrons. We use an efficient on-the-fly Fokker-Planck solver to evolve distributions of CR protons and electrons within every resolution element of our simulation. The solver accounts for CR acceleration at intra-cluster shocks, based on results of recent PIC simulations, re-acceleration due to shocks and MHD turbulence, adiabatic changes and radiative losses of electrons. We apply this model to zoom simulations of galaxy clusters, recently used to show the evolution of the small-scale turbulent dynamo on cluster scales. For these simulations we use a spectral resolution of 48 bins over 6 orders of magnitude in momentum for electrons and 12 bins over 6 orders of magnitude in momentum for protons. We present preliminary results about a possible formation mechanism for \textit{Wrong Way Radio Relics} in our simulation.}

\ConferenceLogo{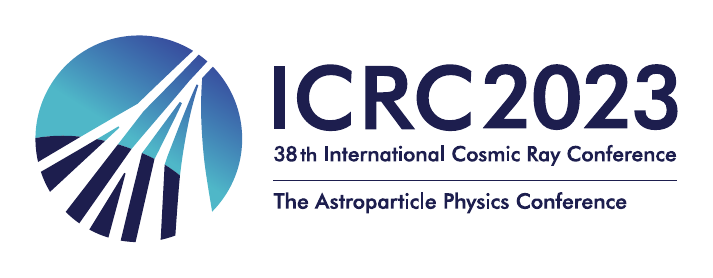}

\FullConference{%
38th International Cosmic Ray Conference (ICRC2023)\\
  26 July - 3 August, 2023\\
  Nagoya, Japan}

\begin{document}
\maketitle

\section{Introduction}

One of the most prominent indications for the presence of Cosmic Ray (CR) electrons in galaxy clusters (GCs) are the so-called "Radio Relics".
These are concave, arc-like structures of up-to several Mpc in size which are found in the periphery of galaxy clusters.
They are typically associated with ongoing mergers between massive galaxy clusters \citep[see e.g.][for reviews]{Brueggen2012, Brunetti2014} and are observed as synchrotron emission in the radio bands where they can show strong polarisation, synchrotron spectral indices typically in the range  $\alpha_0 \sim [-0.6, -0.8]$ and a steepening of these spectra towards the cluster center and with that indicating an aging CR electron population \citep[see][for a recent review]{Weeren2019}.
The common ground in the literature to date is that these relics originate from electrons accelerated at shocks via diffusive shock acceleration (DSA) \citep[e.g.][for a review]{Drury1983}.
However their high surface brightness seems to be incompatible with current models of CR electron acceleration \citep[see e.g.][and references therin]{Botteon2020}, indicating that we lack understanding in the exact process of efficient CR electron acceleration in high-$\beta$ plasmas, with $\beta \equiv \frac{P_\mathrm{th}}{P_B} \sim 20-100$ typical for the intra-cluster medium (ICM).
Recent observations show a class of radio relics that show puzzling features, such as convex morphologies \citep[e.g.][]{Botteon2021, Riseley2022, Rajpurohit2023}, deformations across the shock surface \citep[e.g.][]{deGasperin2022}, filamentary structures \citep[e.g.][]{Rajpurohit2020, Rajpurohit2020a, Rajpurohit2023, Chibueze2023} and strong variations of spectral index across the relic \citep[e.g. the southern relic in][]{DiGennaro2018}.
Simulations can provide insights into some of these features such as complex relic morphology \citep[e.g.][and references therein]{Skillman2013}, surface brightness variations \citep[e.g.][]{Dominguez-Fernandez2021, Wittor2023} and evolution of the spectral index of the relic \citep[e.g.][]{Stroe2016, Donnert2016, Wittor2019} also including multi-shock scenarios \citep[e.g.][]{Inchingolo2022}.
However simulations face the problem that the gap between the mass of an individual electron and the mass of a resolution element typically spans 60-70 orders of magnitude.
Therefore radio emission in simulations are typically studied by applying models for synchrotron emission in post-processing \citep[e.g.][]{Dolag2000, Hoeft2007, Hoeft2008, Vazza2012, Donnert2016, Brunetti2020}, or applying a Fokker-Planck solver to tracer particles to evolve populations of electrons in time \citep[e.g.][]{Donnert2014, Wittor2017, Winner2019}.
We present preliminary results of a sibling-simulation to the one presented in \cite{Steinwandel2023} including the CR model introduced in \cite{Boess2023}, which traces the time evolution of populations of CR electrons and protons within every resolution element of the simulation.

\section{Methods}

The simulation used in this work was carried out with the Tree-SPMHD code \textsc{OpenGadget3}.
\textsc{OpenGadget3} is a derivative of \textsc{Gadget2} \citep{Springel2005} with improvements to the hydro and gravity solvers as well as additional physics modules.
The hydro solver is updated as described in \cite{Beck2016} and we employ the magnetohydrodynamics (MHD) implementation by \cite{Dolag2009} with updates to include non-ideal MHD in the form of constant (physical) diffusion and dissipation, as presented in \cite{Bonafede2011}.
The magnetic dissipation heats the gas with the magnetic field that is lost due to magnetic reconnection.
(An)isotropic conduction is modelled via a (bi-)conjugate gradient solver \citep{Petkova2009, Arth2014, Steinwandel2022}.
We adopt a Wendland C4 kernel with 200 neighbors and bias correction as suggested by \cite{Dehnen2012}.
\\
We employ the on-the-fly spectral CR model \textsc{Crescendo} introduced in \cite{Boess2023} to model the time evolution of CR protons and electrons in every resolution element of our simulation.
The time evolution of distributions of CRs in the absence of CR transport, diffusion in momentum space and catastrophic losses can be described by 
\begin{align}
	\frac{D f(p,\mathbf{x},t)}{Dt} &=  \left( \frac{1}{3} \nabla \cdot \mathbf{u} \right) p \frac{\partial f(p,\mathbf{x},t)}{\partial p} \label{eq:fp-adiabatic} \\
	&+ \frac{1}{p^2} \frac{\partial}{\partial p } \left( p^2 \sum_l b_l f(p,\mathbf{x},t) \right) \label{eq:fp-rad}\\
	&+ j(\mathbf{x}, p, t), \label{eq:fp-sources}
\end{align}
where we used the convective derivative $\frac{Df}{Dt} = \frac{\partial f}{\partial t} + \mathbf{u} \cdot \nabla f$ due to the Lagrangian nature of our code.
The right side of Eq. \ref{eq:fp-adiabatic} describes changes due to adiabatic compression or expansion of the surrounding gas, Eq. \ref{eq:fp-rad} describes energy losses and Eq. \ref{eq:fp-sources} is the source term.
We represent $f(p,\mathbf{x},t)$ as piece-wise powerlaws in momentum space with 2 bins/dex for protons and 8 bins/dex for electrons in the dimensionless momentum range $\hat{p} \equiv \frac{p_i}{m_i c} \in [0.1, 10^5]$, where $p_i$ and $m_i$ refer to the momentum and mass for protons and electrons, respectively.
The distribution function is updated at every timestep following the two-moment approach following \cite{Miniati2001} by computing CR number and energy changes per bin.
We model energy losses of electrons due to synchrotron emission and inverse Compton scattering off CMB photons.
As a source term (Eq.~\ref{eq:fp-sources}) we employ the DSA parametrisation by \cite{Kang2013} for the dependency on sonic Mach number ($\eta(\mathcal{M}_s)$) and the model by \cite{Pais2018} for the dependency on shock obliquity ($\eta(\theta_B)$) and divert a fraction 
\begin{equation}
    \eta_\mathrm{tot} = \eta(\mathcal{M}_s) \times  \eta(\theta_B)
\end{equation}
of the entropy change over the shock into the CR component.
To detect the shock properties on-the-fly in the simulation we use the shock finder introduced by \cite{Beck2016_shock} with improvements to compute the shock obliquity as the angle between the pressure gradient within the kernel (which we treat as the shock normal $\hat{\mathbf{n}}$) and the magnetic field vector upstream of the shock $\mathbf{B}_u$.
The slope of the injected CR spectrum follows linear DSA theory and we use a fixed electron to proton injection ratio of $K_{e/p} = 0.01$.
The CR component exerts feedback on the thermal gas by solving the pressure integral
\begin{equation}
        P_{\mathrm{CR},c} = \frac{1}{3} \: a^{4} \: \int\limits_{p_{\mathrm{min}}}^{p_{\mathrm{cut}}} dp \: 4\pi c p^3 f(p) 
    \label{eq:pressure_integral}
\end{equation}
between the minimum momentum $p_\mathrm{min}$ represented by the CR population and the cutoff of the distribution function $p_\mathrm{cut}$.
We start the CR injection at $z=4$ to avoid too strong time-constraints due to very efficient high-momentum energy losses of CR electrons.
\\
We employ a zoomed-in initial condition of a massive galaxy cluster with a virial mass of $M_\mathrm{vir} \approx 1.3 \times 10^{15}~M_\odot$ from the sample presented in \cite{Bonafede2011}.
The cluster is up-sampled to 250x base resolution, which corresponds to a mass resolution of $M_\mathrm{DM} \approx 4.7 \times 10^6 M_\odot$ and $M_\mathrm{gas} \approx 8.7 \times 10^5 M_\odot$ for dark matter and gas particles, respectively.
This leads to a maximum resolution of $h_\mathrm{sml,min} \approx 1$ kpc.
The parent simulation used a WMAP7 cosmology with $\Omega_{0} = 0.24$, $\Omega_{\Lambda} = 0.76$, $\Omega_\mathrm{baryon} = 0.04$, $h = 0.72$ and $\sigma_{8} = 0.8$, which we also adopt for the present simulation.
We start the simulation at redshift $z=310$ and seed a constant magnetic field in x-direction with $B_0 = $10$^{-14}$ G \citep[see][for a study of the impact of the choice of $B_0$]{Steinwandel2022}.
The initial conditions of this cluster at this resolution have been used to study the interaction between internal- and accretion shocks in \cite{Zhang2020a, Zhang2020b} and its magnetic field has been studied in \cite{Steinwandel2023}.

\section{"Wrong Way" Relics}

\subsection{Merger Geometry}
\begin{figure}
    \centering
    \includegraphics[width=0.9\textwidth]{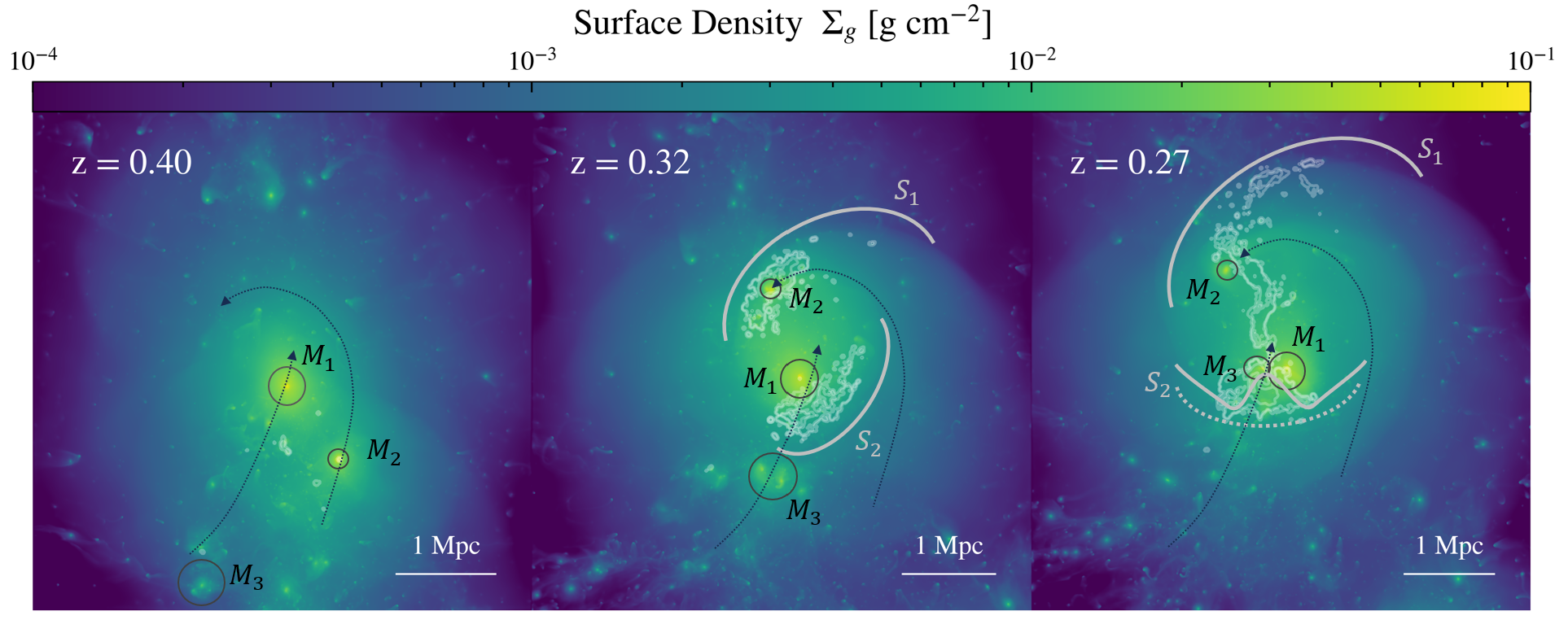}
    \caption{We show the time evolution of the merger geometry from left to right. Arrows indicate the path of the subhalos labeled with $M_1-M_3$. $S_1$ and $S_2$ indicate the resulting merger shocks. Contours show synchrotron surface brightness.}
    \label{fig:merger_schematic}
\end{figure}
The "Wrong Way" relic in our simulation originates from a triple-merger at $z \sim 0.4 - 0.2$.
We show the schematic of the merger geometry in Fig. \ref{fig:merger_schematic}.
A high-velocity merger with a $\sim1:10$ mass ratio between impactor ($M_2$) and target ($M_1$) with a large impact parameter of $b \approx 500$ kpc drives two shock waves.
These shocks follow the canonical picture \citep[e.g. fig. 14 in][]{Weeren2019} of the lighter merging partner ($M_2$) driving a strong bow-shock ($S_1$ in our schematic), while the heavier merging partner ($M_1$) drives a weaker counter shock ($S_2$) in the in-fall direction of the lighter partner.
This counter shock is subsequently impacted by a third merger partner ($M_3$), which ultimately passes through the shock surface and falls into the larger merger partner ($M_1$) in a low-impact parameter merger.
This impact deforms the weaker counter shock ($S_2$) first from a concave shape at $z=0.32$ to a flat shape at $z=0.29$ and subsequently to a convex shape pointing towards the cluster center at $z=0.27$.

\subsection{Time Evolution}
\begin{figure}
    \centering
    \includegraphics[width=0.8\textwidth]{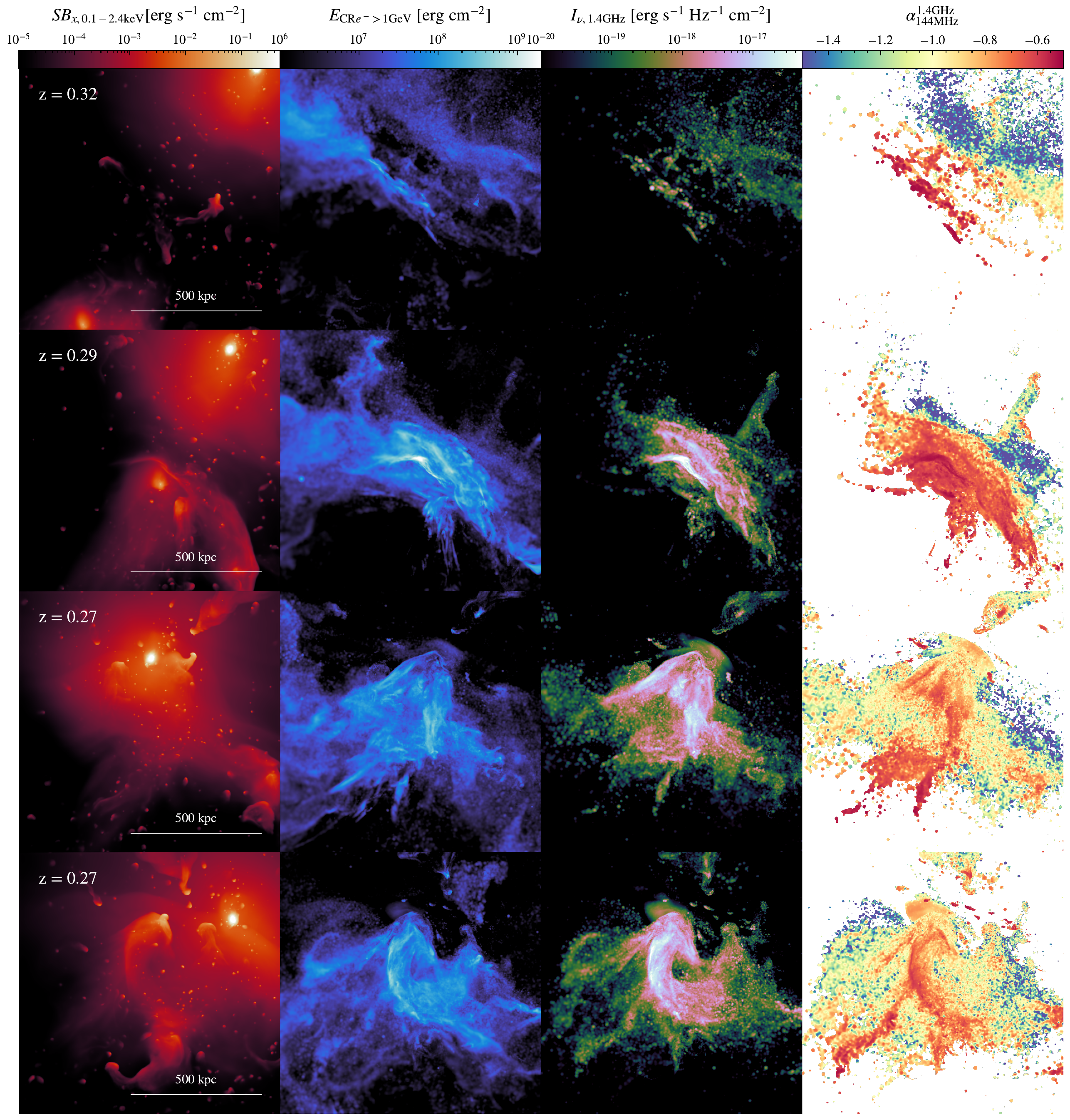}
    \caption{From left to right we show Xray surface brightness, integrated energy density contained in CR electrons with energies above 1 GeV, synchrotron surface brightness at 1.4 GHz and slope of the synchrotron spectrum between 144 MHz and 1.4 GHz. From top to bottom the first three panels show the time evolution of the relic and the lowest panel shows the relic rotated by $90^\circ$.}
    \label{fig:relic}
\end{figure}

Fig. \ref{fig:relic} from top to bottom shows the time evolution of the counter shock $S_2$ and its phasing through morphologies matching various "Wrong Way" relics in the $xz$-plane of the simulation.
The bottom row shows the same relic as the row above in the $yz$-plane.
From left to right we show the Xray surface brightness, CR electron energy contained in the part of the potentially synchrotron bright population with $E > 1$ GeV, the synchrotron surface brightness at 1.4 GHz and the slope of the synchrotron spectrum by fitting a powerlaw to the surface brightness at 144 MHz (not shown) and 1.4 GHz.
Initially, in the top row of Fig. \ref{fig:relic}, we see the acceleration of CR electrons at the counter shock of the main merger event.
We can readily identify the part of the shock surface that accelerates CRs in the center of the images, as it is the most synchrotron bright part and shows a relatively flat synchrotron specrtum.
The steeper parts of the spectrum in the upper right corner of the images indicates that these electrons have been accelerated at ealier times of the shock propagation and have been freely cooling since.
The counter shock is initially not very synchrotron bright, akin to the counter-shock in Abell 3667 \citep[e.g.][]{deGasperin2022}.
The collision between outwards traveling counter shock and the bow-shock of the infalling group ($M_3$) increases the sonic Mach number and with that the acceleration efficiency of the shock, which significantly increases the synchrotron surface brightness at the contact surface of the shocks.
The infalling sub-structure deforms the outwards traveling shock towards a relic pointing "the wrong way", similar to the source \textit{D1} observed by \cite{Riseley2022}.
In the case of our relic the flat spectrum part is further extended, which we attribute to the projection of the relic.
Here the surface of the collision between outwards and inwards traveling shock is pointed roughly $45^{\circ}$ out of the image plane, while \textit{D1} seems to be viewed almost perfectly edge-on.
The collision between shock waves is also visible in our Xray image (left panel, second row in Fig. \ref{fig:relic}), which matches the detection of a shock in Xray by \cite{Sanders2022}.
As the in-falling halo passes through the outwards traveling shock its own bow-shock collides with the older shock, causing the relic to deform further into a v-shaped morphology, such as in the counter shock to the \textit{sausage relic} \citep[e.g.][]{Stroe2013, DiGennaro2018}, or the relic in Abell 2256 \citep[][]{Rajpurohit2023}.
The relic in our simulation shows strong spectral slope variations at $z=0.27$, as can be seen in the right panels of the bottom two rows in Fig. \ref{fig:relic}.
For the lower panels we rotate the image, as this projection more closely resembles the observations of \cite{DiGennaro2018}.
This variation stems from sonic Mach number variations across the surface of the shock.
The majority of the shock surface is super-sonic, but sub-critical ($1 \leq \mathcal{M}_{s} < 2$), leading to a cooling dominated part of the relic surface and an injection dominated part.
The injection dominated part shows a flatter spectrum, indicating that that in our scenario the variations of spectral index can be understood as a superposition of cooling, adiabatic compression and (re-)acceleration.
\section{Discussion}
The shock collision scenario proposed here produces a "Wrong Way" radio relic with convex morphology and a steep synchrotron spectrum as seen in recent observations \citep[e.g.][]{DiGennaro2018, Botteon2021, Riseley2022, Rajpurohit2023}.
The relic also lies within $r_{500}$ which is unlikely in the classical picture of radio relics \citep[e.g.][]{Vazza2012}, indicating that the collision of shocks boosts the (re-)acceleration of CR electrons, making the resulting relics more easy to detect \citep[as suggested in][]{Kang2021}.
The steep spectrum of some of these relics can be explained in a scenario similar to the one here, where electrons accelerated at an outwards traveling shock have already undergone cooling and a collision with an infalling substructure adiabatically compresses the population and some regions experience efficient re-acceleration by the re-energized shock \citep[see e.g.][for a theoretical approach]{Kang2021}.
The strong variations of synchrotron spectral index is still under debate \citep[see discussion in][]{DiGennaro2018}.
In the context of our model it can be understood as a part of the outwards traveling shock being cooling dominated, while the part that collides with an inwards traveling shock is injection dominated.
There are detections of additional infalling substructures along currently ongoing merger axis \citep[e.g.][]{Hoang2019} indicating that the geometric setup is possible, albeit rare.

\section{Conclusion and Outlook}
We showed preliminary results of a high-resolution simulation of a massive galaxy cluster with an on-the-fly Fokker-Planck solver to study the acceleration, advection and aging of CR electrons in a cosmological context.
We used this simulation to study a rare form of radio relics that show convex, instead of the typical concave morphologies.
We find that a multi-merger scenario where the counter-shock of an ongoing merger is deformed by an infalling sub-structure can reproduce the morphologies of some of the current known "Wrong Way" relics.
We will present a more complete study of our proposed formation process of "Wrong Way" Relics in an upcoming publication (B\"oss et. al. (in prep)).
\begin{acknowledgments}
\small
LMB and KD are acknowledging support by the Deutsche Forschungsgemeinschaft (DFG, German Research Foundation) under Germanys Excellence Strategy -- EXC-2094 -- 390783311 and support for the COMPLEX project from the European Research Council (ERC) under the European Union’s Horizon 2020 research and innovation program grant agreement ERC-2019-AdG 860744.
 UPS is supported by the Simons Foundation through a Flatiron Research Fellowship at the Center for Computational Astrophysics of the Flatiron Institute. The Flatiron Institute is supported by the Simons Foundation.
\end{acknowledgments}

\bibliographystyle{mnras}
\bibliography{literature}{}

\begin{thebibliography}{}
\makeatletter
\relax
\def\mn@urlcharsother{\let\do\@makeother \do\$\do\&\do\#\do\^\do\_\do\%\do\~}
\def\mn@doi{\begingroup\mn@urlcharsother \@ifnextchar [ {\mn@doi@}
  {\mn@doi@[]}}
\def\mn@doi@[#1]#2{\def\@tempa{#1}\ifx\@tempa\@empty \href
  {http://dx.doi.org/#2} {doi:#2}\else \href {http://dx.doi.org/#2} {#1}\fi
  \endgroup}
\def\mn@eprint#1#2{\mn@eprint@#1:#2::\@nil}
\def\mn@eprint@arXiv#1{\href {http://arxiv.org/abs/#1} {{\tt arXiv:#1}}}
\def\mn@eprint@dblp#1{\href {http://dblp.uni-trier.de/rec/bibtex/#1.xml}
  {dblp:#1}}
\def\mn@eprint@#1:#2:#3:#4\@nil{\def\@tempa {#1}\def\@tempb {#2}\def\@tempc
  {#3}\ifx \@tempc \@empty \let \@tempc \@tempb \let \@tempb \@tempa \fi \ifx
  \@tempb \@empty \def\@tempb {arXiv}\fi \@ifundefined
  {mn@eprint@\@tempb}{\@tempb:\@tempc}{\expandafter \expandafter \csname
  mn@eprint@\@tempb\endcsname \expandafter{\@tempc}}}

\bibitem[\protect\citeauthoryear{{Arth}, {Dolag}, {Beck}, {Petkova}  \&
  {Lesch}}{{Arth} et~al.}{2014}]{Arth2014}
{Arth} A.,  {Dolag} K.,  {Beck} A.~M.,  {Petkova} M.,   {Lesch} H.,  2014,
  arXiv e-prints, \href {https://ui.adsabs.harvard.edu/abs/2014arXiv1412.6533A}
  {p. arXiv:1412.6533}

\bibitem[\protect\citeauthoryear{{Beck} et~al.,}{{Beck}
  et~al.}{2016a}]{Beck2016}
{Beck} A.~M.,  et~al., 2016a, \mn@doi [MNRAS] {10.1093/mnras/stv2443}, \href
  {https://ui.adsabs.harvard.edu/abs/2016MNRAS.455.2110B} {455, 2110}

\bibitem[\protect\citeauthoryear{{Beck}, {Dolag}  \& {Donnert}}{{Beck}
  et~al.}{2016b}]{Beck2016_shock}
{Beck} A.~M.,  {Dolag} K.,   {Donnert} J.~M.~F.,  2016b, \mn@doi [MNRAS]
  {10.1093/mnras/stw487}, \href
  {https://ui.adsabs.harvard.edu/abs/2016MNRAS.458.2080B} {458, 2080}

\bibitem[\protect\citeauthoryear{{Bonafede}, {Dolag}, {Stasyszyn}, {Murante}
  \& {Borgani}}{{Bonafede} et~al.}{2011}]{Bonafede2011}
{Bonafede} A.,  {Dolag} K.,  {Stasyszyn} F.,  {Murante} G.,   {Borgani} S.,
  2011, \mn@doi [MNRAS] {10.1111/j.1365-2966.2011.19523.x}, \href
  {https://ui.adsabs.harvard.edu/abs/2011MNRAS.418.2234B} {418, 2234}

\bibitem[\protect\citeauthoryear{{B{\"o}ss}, {Steinwandel}, {Dolag}  \&
  {Lesch}}{{B{\"o}ss} et~al.}{2023}]{Boess2023}
{B{\"o}ss} L.~M.,  {Steinwandel} U.~P.,  {Dolag} K.,   {Lesch} H.,  2023,
  \mn@doi [MNRAS] {10.1093/mnras/stac3584}, \href
  {https://ui.adsabs.harvard.edu/abs/2023MNRAS.519..548B} {519, 548}

\bibitem[\protect\citeauthoryear{{Botteon}, {Brunetti}, {Ryu}  \&
  {Roh}}{{Botteon} et~al.}{2020}]{Botteon2020}
{Botteon} A.,  {Brunetti} G.,  {Ryu} D.,   {Roh} S.,  2020, \mn@doi [A&A]
  {10.1051/0004-6361/201936216}, \href
  {https://ui.adsabs.harvard.edu/abs/2020A&A...634A..64B} {634, A64}

\bibitem[\protect\citeauthoryear{{Botteon} et~al.,}{{Botteon}
  et~al.}{2021}]{Botteon2021}
{Botteon} A.,  et~al., 2021, \mn@doi [ApJl] {10.3847/2041-8213/ac0636}, \href
  {https://ui.adsabs.harvard.edu/abs/2021ApJ...914L..29B} {914, L29}

\bibitem[\protect\citeauthoryear{{Br{\"u}ggen}, {Bykov}, {Ryu}  \&
  {R{\"o}ttgering}}{{Br{\"u}ggen} et~al.}{2012}]{Brueggen2012}
{Br{\"u}ggen} M.,  {Bykov} A.,  {Ryu} D.,   {R{\"o}ttgering} H.,  2012, \mn@doi
  [SSR] {10.1007/s11214-011-9785-9}, \href
  {https://ui.adsabs.harvard.edu/abs/2012SSRv..166..187B} {166, 187}

\bibitem[\protect\citeauthoryear{{Brunetti} \& {Jones}}{{Brunetti} \&
  {Jones}}{2014}]{Brunetti2014}
{Brunetti} G.,  {Jones} T.~W.,  2014, \mn@doi [International Journal of Modern
  Physics D] {10.1142/S0218271814300079}, \href
  {https://ui.adsabs.harvard.edu/abs/2014IJMPD..2330007B} {23, 1430007}

\bibitem[\protect\citeauthoryear{{Brunetti} \& {Vazza}}{{Brunetti} \&
  {Vazza}}{2020}]{Brunetti2020}
{Brunetti} G.,  {Vazza} F.,  2020, \mn@doi [PRL]
  {10.1103/PhysRevLett.124.051101}, \href
  {https://ui.adsabs.harvard.edu/abs/2020PhRvL.124e1101B} {124, 051101}

\bibitem[\protect\citeauthoryear{{Chibueze} et~al.,}{{Chibueze}
  et~al.}{2023}]{Chibueze2023}
{Chibueze} J.~O.,  et~al., 2023, \mn@doi [PASJ] {10.1093/pasj/psac009}, \href
  {https://ui.adsabs.harvard.edu/abs/2023PASJ...75S..97C} {75, S97}

\bibitem[\protect\citeauthoryear{{Dehnen} \& {Aly}}{{Dehnen} \&
  {Aly}}{2012}]{Dehnen2012}
{Dehnen} W.,  {Aly} H.,  2012, \mn@doi [MNRAS]
  {10.1111/j.1365-2966.2012.21439.x}, \href
  {https://ui.adsabs.harvard.edu/abs/2012MNRAS.425.1068D} {425, 1068}

\bibitem[\protect\citeauthoryear{{Di Gennaro} et~al.,}{{Di Gennaro}
  et~al.}{2018}]{DiGennaro2018}
{Di Gennaro} G.,  et~al., 2018, \mn@doi [ApJ] {10.3847/1538-4357/aad738}, \href
  {https://ui.adsabs.harvard.edu/abs/2018ApJ...865...24D} {865, 24}

\bibitem[\protect\citeauthoryear{{Dolag} \& {En{\ss}lin}}{{Dolag} \&
  {En{\ss}lin}}{2000}]{Dolag2000}
{Dolag} K.,  {En{\ss}lin} T.~A.,  2000, \mn@doi [A&A]
  {10.48550/arXiv.astro-ph/0008333}, \href
  {https://ui.adsabs.harvard.edu/abs/2000A&A...362..151D} {362, 151}

\bibitem[\protect\citeauthoryear{{Dolag} \& {Stasyszyn}}{{Dolag} \&
  {Stasyszyn}}{2009}]{Dolag2009}
{Dolag} K.,  {Stasyszyn} F.,  2009, \mn@doi [MNRAS]
  {10.1111/j.1365-2966.2009.15181.x}, \href
  {https://ui.adsabs.harvard.edu/abs/2009MNRAS.398.1678D} {398, 1678}

\bibitem[\protect\citeauthoryear{{Dom{\'\i}nguez-Fern{\'a}ndez}, {Br{\"u}ggen},
  {Vazza}, {Banda-Barragan}, {Rajpurohit}, {Mignone}, {Mukherjee}  \&
  {Vaidya}}{{Dom{\'\i}nguez-Fern{\'a}ndez}
  et~al.}{2021}]{Dominguez-Fernandez2021}
{Dom{\'\i}nguez-Fern{\'a}ndez} P.,  {Br{\"u}ggen} M.,  {Vazza} F.,
  {Banda-Barragan} W.~E.,  {Rajpurohit} K.,  {Mignone} A.,  {Mukherjee} D.,
  {Vaidya} B.,  2021, \mn@doi [MNRAS] {10.1093/mnras/staa3018}, \href
  {https://ui.adsabs.harvard.edu/abs/2021MNRAS.500..795D} {500, 795}

\bibitem[\protect\citeauthoryear{{Donnert}}{{Donnert}}{2014}]{Donnert2014}
{Donnert} J.~M.~F.,  2014, \mn@doi [MNRAS] {10.1093/mnras/stt2291}, \href
  {https://ui.adsabs.harvard.edu/abs/2014MNRAS.438.1971D} {438, 1971}

\bibitem[\protect\citeauthoryear{{Donnert}, {Stroe}, {Brunetti}, {Hoang}  \&
  {Roettgering}}{{Donnert} et~al.}{2016}]{Donnert2016}
{Donnert} J.~M.~F.,  {Stroe} A.,  {Brunetti} G.,  {Hoang} D.,   {Roettgering}
  H.,  2016, \mn@doi [MNRAS] {10.1093/mnras/stw1792}, \href
  {https://ui.adsabs.harvard.edu/abs/2016MNRAS.462.2014D} {462, 2014}

\bibitem[\protect\citeauthoryear{Drury}{Drury}{1983}]{Drury1983}
Drury L.~O.,  1983, \mn@doi [Reports on Progress in Physics]
  {10.1088/0034-4885/46/8/002}, 46, 973

\bibitem[\protect\citeauthoryear{{Hoang} et~al.,}{{Hoang}
  et~al.}{2019}]{Hoang2019}
{Hoang} D.~N.,  et~al., 2019, \mn@doi [A&A] {10.1051/0004-6361/201833900},
  \href {https://ui.adsabs.harvard.edu/abs/2019A&A...622A..20H} {622, A20}

\bibitem[\protect\citeauthoryear{{Hoeft} \& {Br{\"u}ggen}}{{Hoeft} \&
  {Br{\"u}ggen}}{2007}]{Hoeft2007}
{Hoeft} M.,  {Br{\"u}ggen} M.,  2007, \mn@doi [MNRAS]
  {10.1111/j.1365-2966.2006.11111.x}, \href
  {https://ui.adsabs.harvard.edu/abs/2007MNRAS.375...77H} {375, 77}

\bibitem[\protect\citeauthoryear{{Hoeft}, {Br{\"u}ggen}, {Yepes},
  {Gottl{\"o}ber}  \& {Schwope}}{{Hoeft} et~al.}{2008}]{Hoeft2008}
{Hoeft} M.,  {Br{\"u}ggen} M.,  {Yepes} G.,  {Gottl{\"o}ber} S.,   {Schwope}
  A.,  2008, \mn@doi [MNRAS] {10.1111/j.1365-2966.2008.13955.x}, \href
  {https://ui.adsabs.harvard.edu/abs/2008MNRAS.391.1511H} {391, 1511}

\bibitem[\protect\citeauthoryear{{Inchingolo}, {Wittor}, {Rajpurohit}  \&
  {Vazza}}{{Inchingolo} et~al.}{2022}]{Inchingolo2022}
{Inchingolo} G.,  {Wittor} D.,  {Rajpurohit} K.,   {Vazza} F.,  2022, \mn@doi
  [MNRAS] {10.1093/mnras/stab3096}, \href
  {https://ui.adsabs.harvard.edu/abs/2022MNRAS.509.1160I} {509, 1160}

\bibitem[\protect\citeauthoryear{{Kang}}{{Kang}}{2021}]{Kang2021}
{Kang} H.,  2021, \mn@doi [arXiv e-prints] {10.48550/arXiv.2108.01876}, \href
  {https://ui.adsabs.harvard.edu/abs/2021arXiv210801876K} {p. arXiv:2108.01876}

\bibitem[\protect\citeauthoryear{{Kang} \& {Ryu}}{{Kang} \&
  {Ryu}}{2013}]{Kang2013}
{Kang} H.,  {Ryu} D.,  2013, \mn@doi [ApJ] {10.1088/0004-637X/764/1/95}, \href
  {https://ui.adsabs.harvard.edu/abs/2013ApJ...764...95K} {764, 95}

\bibitem[\protect\citeauthoryear{Miniati}{Miniati}{2001}]{Miniati2001}
Miniati F.,  2001, \mn@doi [Computer Physics Communications]
  {10.1016/s0010-4655(01)00293-4}, 141, 17

\bibitem[\protect\citeauthoryear{{Pais}, {Pfrommer}, {Ehlert}  \&
  {Pakmor}}{{Pais} et~al.}{2018}]{Pais2018}
{Pais} M.,  {Pfrommer} C.,  {Ehlert} K.,   {Pakmor} R.,  2018, \mn@doi [MNRAS]
  {10.1093/mnras/sty1410}, \href
  {https://ui.adsabs.harvard.edu/abs/2018MNRAS.478.5278P} {478, 5278}

\bibitem[\protect\citeauthoryear{{Petkova} \& {Springel}}{{Petkova} \&
  {Springel}}{2009}]{Petkova2009}
{Petkova} M.,  {Springel} V.,  2009, \mn@doi [MNRAS]
  {10.1111/j.1365-2966.2009.14843.x}, \href
  {https://ui.adsabs.harvard.edu/abs/2009MNRAS.396.1383P} {396, 1383}

\bibitem[\protect\citeauthoryear{{Rajpurohit} et~al.,}{{Rajpurohit}
  et~al.}{2020a}]{Rajpurohit2020}
{Rajpurohit} K.,  et~al., 2020a, \mn@doi [A&A] {10.1051/0004-6361/201937139},
  \href {https://ui.adsabs.harvard.edu/abs/2020A&A...636A..30R} {636, A30}

\bibitem[\protect\citeauthoryear{{Rajpurohit} et~al.,}{{Rajpurohit}
  et~al.}{2020b}]{Rajpurohit2020a}
{Rajpurohit} K.,  et~al., 2020b, \mn@doi [A&A] {10.1051/0004-6361/202039165},
  \href {https://ui.adsabs.harvard.edu/abs/2020A&A...642L..13R} {642, L13}

\bibitem[\protect\citeauthoryear{{Rajpurohit} et~al.,}{{Rajpurohit}
  et~al.}{2023}]{Rajpurohit2023}
{Rajpurohit} K.,  et~al., 2023, \mn@doi [A&A] {10.1051/0004-6361/202244925},
  \href {https://ui.adsabs.harvard.edu/abs/2023A&A...669A...1R} {669, A1}

\bibitem[\protect\citeauthoryear{{Riseley} et~al.,}{{Riseley}
  et~al.}{2022}]{Riseley2022}
{Riseley} C.~J.,  et~al., 2022, \mn@doi [MNRAS] {10.1093/mnras/stac1771}, \href
  {https://ui.adsabs.harvard.edu/abs/2022MNRAS.515.1871R} {515, 1871}

\bibitem[\protect\citeauthoryear{{Sanders} et~al.,}{{Sanders}
  et~al.}{2022}]{Sanders2022}
{Sanders} J.~S.,  et~al., 2022, \mn@doi [A&A] {10.1051/0004-6361/202141501},
  \href {https://ui.adsabs.harvard.edu/abs/2022A&A...661A..36S} {661, A36}

\bibitem[\protect\citeauthoryear{{Skillman}, {Xu}, {Hallman}, {O'Shea},
  {Burns}, {Li}, {Collins}  \& {Norman}}{{Skillman}
  et~al.}{2013}]{Skillman2013}
{Skillman} S.~W.,  {Xu} H.,  {Hallman} E.~J.,  {O'Shea} B.~W.,  {Burns} J.~O.,
  {Li} H.,  {Collins} D.~C.,   {Norman} M.~L.,  2013, \mn@doi [ApJ]
  {10.1088/0004-637X/765/1/21}, \href
  {https://ui.adsabs.harvard.edu/abs/2013ApJ...765...21S} {765, 21}

\bibitem[\protect\citeauthoryear{{Springel}}{{Springel}}{2005}]{Springel2005}
{Springel} V.,  2005, \mn@doi [MNRAS] {10.1111/j.1365-2966.2005.09655.x}, \href
  {https://ui.adsabs.harvard.edu/abs/2005MNRAS.364.1105S} {364, 1105}

\bibitem[\protect\citeauthoryear{{Steinwandel}, {B{\"o}ss}, {Dolag}  \&
  {Lesch}}{{Steinwandel} et~al.}{2022}]{Steinwandel2022}
{Steinwandel} U.~P.,  {B{\"o}ss} L.~M.,  {Dolag} K.,   {Lesch} H.,  2022,
  \mn@doi [ApJ] {10.3847/1538-4357/ac715c}, \href
  {https://ui.adsabs.harvard.edu/abs/2022ApJ...933..131S} {933, 131}

\bibitem[\protect\citeauthoryear{{Steinwandel}, {Dolag}, {B{\"o}ss}  \&
  {Marin-Gilabert}}{{Steinwandel} et~al.}{2023}]{Steinwandel2023}
{Steinwandel} U.~P.,  {Dolag} K.,  {B{\"o}ss} L.,   {Marin-Gilabert} T.,  2023,
  \mn@doi [arXiv e-prints] {10.48550/arXiv.2306.04692}, \href
  {https://ui.adsabs.harvard.edu/abs/2023arXiv230604692S} {p. arXiv:2306.04692}

\bibitem[\protect\citeauthoryear{{Stroe}, {van Weeren}, {Intema},
  {R{\"o}ttgering}, {Br{\"u}ggen}  \& {Hoeft}}{{Stroe}
  et~al.}{2013}]{Stroe2013}
{Stroe} A.,  {van Weeren} R.~J.,  {Intema} H.~T.,  {R{\"o}ttgering} H.~J.~A.,
  {Br{\"u}ggen} M.,   {Hoeft} M.,  2013, \mn@doi [A&A]
  {10.1051/0004-6361/201321267}, \href
  {https://ui.adsabs.harvard.edu/abs/2013A&A...555A.110S} {555, A110}

\bibitem[\protect\citeauthoryear{{Stroe} et~al.,}{{Stroe}
  et~al.}{2016}]{Stroe2016}
{Stroe} A.,  et~al., 2016, \mn@doi [MNRAS] {10.1093/mnras/stv2472}, \href
  {https://ui.adsabs.harvard.edu/abs/2016MNRAS.455.2402S} {455, 2402}

\bibitem[\protect\citeauthoryear{{Vazza}, {Br{\"u}ggen}, {Gheller}  \&
  {Brunetti}}{{Vazza} et~al.}{2012}]{Vazza2012}
{Vazza} F.,  {Br{\"u}ggen} M.,  {Gheller} C.,   {Brunetti} G.,  2012, \mn@doi
  [MNRAS] {10.1111/j.1365-2966.2012.20562.x}, \href
  {https://ui.adsabs.harvard.edu/abs/2012MNRAS.421.3375V} {421, 3375}

\bibitem[\protect\citeauthoryear{{Winner}, {Pfrommer}, {Girichidis}  \&
  {Pakmor}}{{Winner} et~al.}{2019}]{Winner2019}
{Winner} G.,  {Pfrommer} C.,  {Girichidis} P.,   {Pakmor} R.,  2019, \mn@doi
  [MNRAS] {10.1093/mnras/stz1792}, \href
  {https://ui.adsabs.harvard.edu/abs/2019MNRAS.488.2235W} {488, 2235}

\bibitem[\protect\citeauthoryear{{Wittor}, {Vazza}  \& {Br{\"u}ggen}}{{Wittor}
  et~al.}{2017}]{Wittor2017}
{Wittor} D.,  {Vazza} F.,   {Br{\"u}ggen} M.,  2017, \mn@doi [MNRAS]
  {10.1093/mnras/stw2631}, \href
  {https://ui.adsabs.harvard.edu/abs/2017MNRAS.464.4448W} {464, 4448}

\bibitem[\protect\citeauthoryear{{Wittor}, {Hoeft}, {Vazza}, {Br{\"u}ggen}  \&
  {Dom{\'\i}nguez-Fern{\'a}ndez}}{{Wittor} et~al.}{2019}]{Wittor2019}
{Wittor} D.,  {Hoeft} M.,  {Vazza} F.,  {Br{\"u}ggen} M.,
  {Dom{\'\i}nguez-Fern{\'a}ndez} P.,  2019, \mn@doi [MNRAS]
  {10.1093/mnras/stz2715}, \href
  {https://ui.adsabs.harvard.edu/abs/2019MNRAS.490.3987W} {490, 3987}

\bibitem[\protect\citeauthoryear{{Wittor}, {Br{\"u}ggen}, {Grete}  \&
  {Rajpurohit}}{{Wittor} et~al.}{2023}]{Wittor2023}
{Wittor} D.,  {Br{\"u}ggen} M.,  {Grete} P.,   {Rajpurohit} K.,  2023, \mn@doi
  [MNRAS] {10.1093/mnras/stad1463}, \href
  {https://ui.adsabs.harvard.edu/abs/2023MNRAS.523..701W} {523, 701}

\bibitem[\protect\citeauthoryear{{Zhang}, {Churazov}, {Dolag}, {Forman}  \&
  {Zhuravleva}}{{Zhang} et~al.}{2020a}]{Zhang2020b}
{Zhang} C.,  {Churazov} E.,  {Dolag} K.,  {Forman} W.~R.,   {Zhuravleva} I.,
  2020a, \mn@doi [MNRAS] {10.1093/mnras/staa1013}, \href
  {https://ui.adsabs.harvard.edu/abs/2020MNRAS.494.4539Z} {494, 4539}

\bibitem[\protect\citeauthoryear{{Zhang}, {Churazov}, {Dolag}, {Forman}  \&
  {Zhuravleva}}{{Zhang} et~al.}{2020b}]{Zhang2020a}
{Zhang} C.,  {Churazov} E.,  {Dolag} K.,  {Forman} W.~R.,   {Zhuravleva} I.,
  2020b, \mn@doi [MNRAS] {10.1093/mnrasl/slaa147}, \href
  {https://ui.adsabs.harvard.edu/abs/2020MNRAS.498L.130Z} {498, L130}

\bibitem[\protect\citeauthoryear{{de Gasperin} et~al.,}{{de Gasperin}
  et~al.}{2022}]{deGasperin2022}
{de Gasperin} F.,  et~al., 2022, \mn@doi [A&A] {10.1051/0004-6361/202142658},
  \href {https://ui.adsabs.harvard.edu/abs/2022A&A...659A.146D} {659, A146}

\bibitem[\protect\citeauthoryear{van Weeren, de Gasperin, Akamatsu, Brüggen,
  Feretti, Kang, Stroe  \& Zandanel}{van Weeren et~al.}{2019}]{Weeren2019}
van Weeren R.~J.,  de Gasperin F.,  Akamatsu H.,  Brüggen M.,  Feretti L.,
  Kang H.,  Stroe A.,   Zandanel F.,  2019, \mn@doi [Space Science Reviews]
  {10.1007/s11214-019-0584-z}, 215

\makeatother
\end{thebibliography}



%
%
%

\end{document}